# EXCLUSIVE SEMILEPTONIC AND RARE RADIATIVE B DECAYS FROM QCD SUM RULES

V.M. BRAUN

*NORDITA, Blegdamsvej 17, DK-2100 Copenhagen, Denmark*

I give an overview of the QCD sum rule predictions for the form factors of B-meson weak decays to light hadrons (and leptons).

## 1 Introduction

In this talk I give an overview of the existing QCD sum rule calculations of the form factors in B-meson weak decays, involving the light hadron (meson) in the final state. Description of these decays presents a considerable challenge for the theory and is more involved than, say, of $B \to D, D^* e\nu$ form factors. I will discuss the semileptonic form factors $B \to \pi e\nu$ and $B \to \rho e\nu$ defined as

$$\begin{aligned}
\langle \pi | \bar{q} \Gamma_\mu b | B \rangle &= f_+(q^2)(p_B + p_\pi)_\mu \\
&\quad + f_-(q^2)(p_B - p_\pi)_\mu , \quad (1)
\end{aligned}$$

$$\begin{aligned}
\langle \rho, \lambda | \bar{q} \Gamma_\mu b | B \rangle &= -i(m_B + m_\rho) A_1(q^2) \epsilon_\mu^{(\lambda)} \\
&\quad + \frac{i A_2(q^2)}{m_B + m_\rho}(\epsilon^{(\lambda)} \cdot p_B)(p_B + p_\rho)_\mu \\
&\quad + \frac{i A_3(q^2)}{m_B + m_\rho}(\epsilon^{(\lambda)} \cdot p_B)(p_B - p_\rho)_\mu \\
&\quad + \frac{2V(q^2)}{m_B + m_\rho} \varepsilon_{\mu\nu\alpha\beta} \epsilon^{\nu,(\lambda)} p_B^\alpha p_\rho^\beta \quad (2)
\end{aligned}$$

In fact, $f_-(q^2)$ and $A_3(q^2)$ do not contribute to the decay rate and will be omitted. In addition, I will summarize the existing results for the rare radiative form factors (short-distance) $B \to V + \gamma$, where $V = K^*, \rho, \omega$

$$\begin{aligned}
\langle V, \lambda | \bar{q} \sigma_{\mu\nu} q^\nu (1+\gamma_5) b | B \rangle &= \\
= \left[ \epsilon_\mu^{(\lambda)}(q \cdot p_B) - p_{B,\mu}(q \cdot \epsilon^{(\lambda)}) \right] &\cdot 2 F_1^S(q^2) \\
+ i \varepsilon_{\mu\nu\alpha\beta} \epsilon^{(\lambda)\nu} p_B^\alpha q^\beta &\cdot 2 F_2^S(q^2) , \quad (3)
\end{aligned}$$

and very briefly discuss the estimates of long-distance effects in these decays, induced by four-fermion operators.

## 2 Theoretical Status

A common feature of the "heavy-to-light" decays is that they involve a large interval of possible invariant energy transfer $0 < q^2 < (m_B - m_{\pi,\rho,...})^2 \simeq m_b^2$, where $m_b$ is the b-quark mass. The region of small recoil $m_b^2 - q^2 \sim O(m_b)$ is simpler, since the light quark produced in the decay is soft, and one can apply the heavy quark expansion techniques. The spin-flavor symmetry in the $m_b \to \infty$ limit induces important relations between the radiative and the semileptonic form factors derived by Isgur and Wise[1]. The pole dominance approximation is expected to become exact at small recoil in this limit[2]. This region is also easier to treat within quark models.

The major part of the decay rate comes, however, from the region $m_b^2 - q^2 \sim O(m_b^2)$ in which case the physics is different. The recoiling light quark carries large energy of order $m_b/2$ and has to transfer it to the soft cloud to recombine in the final state hadron. This question — how to transfer a large momentum to a hadron — has attracted quite a bit of theoretical attention in the past in connection with form factors of light hadrons, see[3] for a review and references to original research. One possibility, called the hard rescattering mechanism, is to find the hadron in the configuration with a minimum number of Fock constituents at small transverse separation, and exchange a hard gluon. Another option is to pick up the configuration in which the active quark carries almost all the momentum of the hadron, so that the recombination with soft spectators does not involve any hard exchanges and relevant transverse distances can be large. This is called the soft contribution, or the Feynman mechanism. The result of crucial importance is that meson wave functions (distribution amplitudes in the fraction $x$ of the total momentum carried by one quark) behave at $x \to 1$ as $\phi(x) \sim 1 - x$. This implies that the soft contribution is of order $1/Q^4$ and is power suppressed compared to the hard rescattering which provides a contribution to the form factor of order $1/Q^2$.

The situation with the decay form factors at large recoil is quite different. Indeed, it turns out[9] that in the "heavy-to-light" decays both the "soft" and "hard" contributions have the *same power behaviour* at $m \to \infty$. At $q^2 = 0$ both contributions are of order $1/m^{3/2}$ for all the form factors defined above. In fact, at extremely large $m_b$ the contribution coming from large transverse distances is suppressed by the Sudakov form factor, see[4] for details. Hence the hard rescattering mechanism dominates asymptotically also in this case. However, at $m_b \sim 5$ GeV the Sudakov suppression is still weak and the soft contribution is more important because it does not involve a small factor $\alpha_s(m_b)/\pi$. Dominance of the soft contribution in B-decays is supported by two kind of evidence: First, the direct calculations of the hard contributions[4] give results much below current model estimates; Second, it is becoming increasingly clear that the soft contribution is important for the description of the pion form factor up to large $Q^2 \sim 10$ GeV$^2$, despite its additional suppression in this case by a power of $1/Q^2$. (Sudakov suppression is common for both cases, although the formalism is somewhat different.)

## 3  QCD sum rules

Since the soft contribution to the decay form factors involves contribution of large transverse separations, its evaluation requires a certain nonperturbative technique. The QCD sum rule approach is due to Shifman, Vainshtein and Zakharov[5]. First QCD sum rule calculations of form factors have been done[6] for light quarks. Extension of this method to heavy hadron decays is straightforward at the point of maixmum recoil $q^2 = 0$, but has some difficulties for $q^2 > 0$. This problem was solved in[7], which made possible to make predictions for the $q^2$-dependence of the weak decay form factors. At present there exist quite a few calculations done in this "traditional" framework. It is worth while to note that contributions of hard rescattering correspond to radiative corrections to the sum rules; they can be included in principle, but this was not done so far.

In addition, a modification of the QCD sum rule approach has been developed, usually referred to as "light-cone sum rules". This approach combines the QCD sum rules techniques with the information about light-cone hadron wave functions available from the theory of exclusive processes. This method was suggested[8] initially for the study of the weak radiative decay $\Sigma \to p\gamma$. Chernyak and Zhitnitsky[9] were first to realize the potential of this approach for the heavy hadron decays. The term "light-cone sum rules" first appears in [7].

Being similar in spirit, these two approaches differ in the treatment of the light hadron in the final state. This is illustrated in Fig. 1. As mentioned above, the form factor is essentially given by the overlap integral of the hadron wave function with a very asymmetric quark-antiquark state, with almost all the momentum carried by the quark, see Fig. 1a: $1-x \sim O(1/m_b)$. Since in QCD the corresponding amplitude behaves as $\phi(x) \sim 1-x$, the form factor is of order $\int_{1-O(1/m_b)}^{1} dx\, \phi(x) \sim O(1/m_b^2)$. In fact, an additional factor $m_b^{1/2}$ comes from the normalization of the heavy initial state, so that the final scaling law is $1/m_b^{3/2}$[9,16,4]. The traditional sum rules avoid introduction of the wave function by considering the correlation function with a suitable interpolating current and using dispersion relations to extract the contribution of the ground state. The most important nonperturbative effect is then described by the diagram as in Fig. 1b, with the light quark put in the condensate (shown by crosses). Since quarks in the condensate have zero momentum, it is easy to see that this contribution is naively proportional to $\delta(1-x)$, and remains unsuppressed when $m_b \to \infty$. The controversy must of course be resolved by higher-order condensate contributions to the sum rules, and subtraction of the contribution of excited states. Thus, the suppression of the end-point region is expected to hold in the sum rules as a numerical cancellation between different contributions, and this cancellation becomes more delicate as $m_b$ increases.

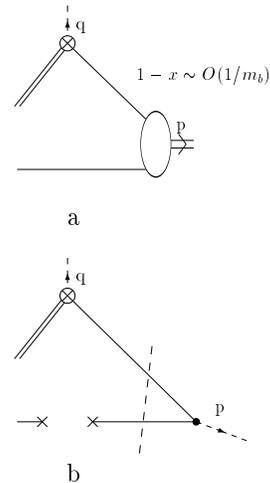

Figure 1: The soft contribution to the decay form factor (a) and its modelling in the QCD sum rule approach (b)

The light-cone sum rules avoid this problem by rearranging the calculation in such a way that the nonperturbative effects like the interaction with the quark condensate are included in the *nonperturbative* hadron distribution amplitudes, estimated using additional sum rules[10]. These additional sum rules are written for integrated characteristics of the wave functions like moments, and the correct asymptotic behaviour at the end points is assumed. It should be added that "traditional" nonperturbative corrections are in general distributed among wave functions of different twist.

The premium is that the light-cone sum rules have the correct asymptotic behaviour in the heavy quark limit, but the problem is that the present knowledge of higher-twist wave functions is incomplete so that not all known nonperturbative corrections of the standard approach can be included. Since the b-quark mass is not that large, one should expect that these two approaches provide with complimentary descriptions for B-decays, with their own advantages and disadvantages. The spread of results can be considered as a theoretical error, borrowing in mind that the numbers coming from traditional sum rules are more or less final, while the light-cone results can still be improved.

Table 1: QCD sum rule predictions for semileptonic $b \to u$ form factors at $q^2 = 0$.

| Ref. | $f_+^{B\to\pi}$ | $A_1^{B\to\rho}$ | $A_2^{B\to\rho}$ | $V^{B\to\rho}$ |
|---|---|---|---|---|
| 11 | 0.4±0.1 | – | – | – |
| 9  | 0.36 | – | – | – |
| 12 | 0.24±0.025 | – | – | – |
| 13 | 0.23±0.02 | 0.35±0.16 | 0.42±0.12 | 0.47±0.14 |
| 14 | 0.26±0.02 | 0.5±0.1 | 0.4±0.2 | 0.6±0.2 |
| 15 | 0.24–0.29 | – | – | – |
| 16 | – | 0.24±0.04 | – | 0.28±0.06 |

## 4 Results

### 4.1 Form Factors at $q^2 = 0$

The form factors at maximum recoil $q^2 = 0$ are most important for phenomenology. The results for semileptonic and rare radiative form factors are collected in Table 1 and Table 2, respectively. Since $F_1(q^2 = 0) = F_2(q^2 = 0)$, see (3), I drop the subscript. The calculations in [9,15,16] are done using light-cone sum rules, the other ones use the traditional technique. There is a general agreement for $B \to \pi e\nu$ semileptonic decays, with the result $f_+(0) \simeq 0.25 - 0.30$. Similarly, all authors give $F(0) = 0.3 - 0.35$ for the radiative form factor $B \to K^*\gamma$ which is, however, somewhat misleading, since different input parameters are used. The light-cone sum rules generally yield somewhat lower values compared to traditional sum rules. The situation is not clear at the moment for $B \to \rho e\nu$ decays, where their is a disagreement between the light-cone sum rules[16] and traditional sum rules[14] by roughly factor two. It is possible to check[7] that the same sum rules agree for D-decays, so that one can suspect traditional approach in overestimating the end-point contribution with a larger b-quark mass. Additional study is necessary to clarify this issue[22].

The heavy quark mass dependence of form factors is of considerable interest for lattice calculations, where the data are collected mostly in the D-meson range. The $m_b$ dependence of the sum rule results for $F(0)$ was analysed in some detail in[16], with the result

$$F_1^{B\to K^*\gamma} \sim 1/m_b \qquad (4)$$

in the interval of quark masses between 1.5 and 5 GeV. This behaviour holds numerically in the specified interval of masses and should not be confused with the asymptotic behaviour discussed above.

It was argued[23] that the Isgur-Wise relations between radiative and semileptonic form factors have a chance to hold with good accuracy over the whole region of $q^2$. This was checked in the light-cone approach[16], and indeed a good agreement was found. Usefullness of these relations can, however, be limited by $SU(3)$ breaking corrections which can be large if the form factors are dominated by the end-point soft contribution. Indeed, the effect of a small spectator quark mass is more pronounced in the end-point region than on the average. The estimates are[16]

$$\begin{aligned} A_1(0)^{B\to\rho}/A_1(0)^{B\to K^*} &= 0.76 \pm 0.05 \,, \\ V(0)^{B\to\rho}/V(0)^{B\to K^*} &= 0.73 \pm 0.05 \,. \end{aligned} \qquad (5)$$

The traditional sum rules suggest a somewhat smaller $SU(3)$ breaking of order 15%[21].

### 4.2 The $q^2$-dependence

The QCD sum rules were probably the first to predict the $q^2$ behaviour of the weak decay form factors, instead of assuming a certain form, as common in quark models. General trends found in the first calculations[7,12] were later confirmed by later analysis.

First result[12] was an (unexpected) approximate pole dominance behaviour of the $B \to \pi e\nu$ form factor in a wide interval of $q^2$, confirmed by the light-cone sum rule calculation in[15]. The vector dominance approximation predicts not only the shape, but also the normalization of the form factor, which is governed by the coupling $g_{BB^*\pi}$. This coupling was calculated using similar techniques (see[24] and references therein) and it was checked that the value of $f_+(0)$ corresponding to would-be-exact vector dominance is only about 30% higher that the result of the direct calculation. On the evidence of existing calculations[7,14,16] approximate pole dominance behaviour is expected also for the semileptonic form factor $V(q^2)$ and for the form factors of rare decays.

Second, it was found[7,14] that axial form factors not at all obey the pattern of pole dominance. In particular, the semileptonic form factor $A_1$ comes out to be much more flat. The traditional sum rules[7,14] even suggest a decreasing form factor at large $q^2$, while the light-cone sum rules[16] still indicate a moderate increase, less steep than for vector form factors. This different behaviour of axial and vector form factors can have important effect on measured asymmetries.

Table 2: QCD sum rule predictions for rare radiative B-decay form factors at $q^2 = 0$.

| Ref. | $F^{B \to K^* \gamma}$ | $F^{B_u \to \rho \gamma}$ | $F^{B_s \to \phi \gamma}$ | $F^{B_s \to K^* \gamma}$ |
|------|------------------------|---------------------------|---------------------------|--------------------------|
| 17   | 0.5±0.1                | –                         | –                         | –                        |
| 18   | 0.56±0.10              | –                         | –                         | –                        |
| 19   | 0.38±0.05              | –                         | –                         | –                        |
| 20   | 0.35±0.05              | –                         | –                         | –                        |
| 16   | 0.32±0.05              | 0.24±0.04                 | 0.29±0.05                 | 0.20±0.04                |
| 21   | 0.31±0.04              | 0.27±0.04                 | –                         | –                        |

### 4.3 Long-distance effects in Radiative B-Decays

Rare radiative B-decays are mainly interesting as a source of information about top quark physics encoded in contributions of penguin operators. Several authors raised a question whether these contributions will not be obscured by long-distance effects induced by four-fermion operators. Very recently, the contribution of weak annihilation was estimated in [25,26] using light-cone sum rules. For the ratio of the long-distance to the short distance amplitudes in the $B_u \to \rho^+ \gamma$ decays it was found

$$|A_{\text{long}}/A_{\text{short}}|^{B_u \to \rho^+ \gamma} = |R_{L/S}| \left| \frac{V_{ub} V_{ud}}{V_{td} V_{tb}} \right| \quad (6)$$

with $R_{L/S} = -0.30 \pm 0.07$ [26], which is a 10% effect. The weak annihilation contributions to the neutral B-decays are expected to be much smaller since they are colour-suppressed. Using this, one may try to determine the sign of the Wolfenstein $\rho$ parameter by measuring the ratio of the decay rates

$$\frac{\Gamma(B_u \to \rho \gamma)}{2\Gamma(B_d \to \rho \gamma)} = 1 + 2 \cdot R_{L/S} V_{ud} \frac{\rho(1-\rho) - \eta^2}{(1-\rho)^2 + \eta^2}$$
$$+ (R_{L/S})^2 V_{ud}^2 \frac{\rho^2 + \eta^2}{(1-\rho)^2 + \eta^2} \quad (7)$$

which should be larger (smaller) than unity depending on $\rho$ being negative (positive). The effect of weak annihilation onto the ratio $\Gamma(B_u \to \rho \gamma)/\Gamma(B \to K^* \gamma)$ (cf.[16]) is smaller than present uncertainty in the SU(3) breaking.

### Acknowledgments

I am grateful to A. Ali, I. Balitsky, P. Ball, V.M. Belyaev, H.G. Dosch, A. Khodjamiryan, R. Rückl, and H. Simma for the collaboration on subjects related to this report.